\documentclass[final,english]{bullsrsl}

\usepackage[latin1]{inputenc}

\usepackage[T1]{fontenc}

\usepackage{natbib} 

\usepackage{graphicx}

\begin{document}

\title{Overview And Present Status Of The 4-m International Liquid Mirror Telescope (ILMT)}



  \author[affil={1,2}, corresponding]{Jean}{Surdej}
  \author[affil={3}]{Bhavya}{Ailawadhi}
  \author[affil={4,5}]{Talat}{Akhunov}
  \author[affil={2,6}]{Monalisa}{Dubey}
  \author[affil={2,6}]{Naveen}{Dukiya}  
  \author[affil={1}]{Sara}{Filali}
  \author[affil={7,8}]{Paul}{Hickson}  
  \author[affil={2}]{Priyanshi}{Kumari}
  \author[affil={2}]{Gokul Singh}{Mehra}
  \author[affil={2}]{Kuntal}{Misra}
  \author[affil={9}]{Vibore}{Negi}  
  \author[affil={1}]{Anna}{Pospieszalska-Surdej}
  \author[affil={2,10}]{Kumar}{Pranshu}  
  \author[affil={2,10}]{Sarvesh}{Yadav}

\affiliation[1]{Institute of Astrophysics and Geophysics, Li\`ege University, All\'ee du 6 Ao\^ut 19c, 4000 Li\`ege, Belgium}
\affiliation[2]{Aryabhatta Research Institute of Observational Sciences, Manora Peak, Nainital, 263001 Uttarakhand, India}
\affiliation[3]{Physical Research Laboratory, Ahmedabad 380009, Gujarat, India}
\affiliation[4]{National University of Uzbekistan, Department of Astronomy and Astrophysics, 100174 Tashkent, Uzbekistan}
\affiliation[5]{Ulugh Beg Astronomical Institute of the Uzbek Academy of Sciences, Astronomicheskaya 33, 100052 Tashkent, Uzbekistan}
\affiliation[6]{Mahatma Jyotiba Phule Rohilkhand University, Pilibhit Bypass Road, Bareilly 243006, Uttar Pradesh, India}
\affiliation[7]{Department of Physics and Astronomy, The University of British Columbia, 6224 Agricultural Road, Vancouver, BC V6T 1Z1,
Canada}
\affiliation[8]{Outer Space Institute, The University of British Columbia, 325-6224 Agricultural Road, Vancouver, BC V6T 1Z1, Canada}
\affiliation[9]{Kavli Institute for Astronomy and Astrophysics, Peking University, Beijing 100871, China}
\affiliation[10]{University of Calcutta, 87/1 College Street, Kolkata 700073, India}


\correspondance{jsurdej@uliege.be}


\maketitle

\begin{abstract}
The International Liquid Mirror Telescope (ILMT) is an inexpensive optical instrument with a diameter of 4 metres, entirely dedicated to astronomical surveys. Its scientific programmes include the detection and study of transients, variable objects, asteroids, comets, low surface brightness galaxies, space debris, etc. The telescope achieved its first light on 29 April 2022 and was officially inaugurated on 21 March 2023. It has been operating regularly since October 2023 and continuously observes the sky passing through the zenith using the SDSS $g^{\prime}$, $r^{\prime}$, and $i^{\prime}$ filters in a narrow band of the sky ($22.3^{\prime}$) centred around the equatorial declination of
$+29^\circ 21^{\prime}41^{\prime\prime}$. Here we report on the astrometric and photometric performance of the ILMT and summarise the main results obtained since the BINA3 workshop.  A great way to raise public awareness is to build a scale model of a liquid mirror telescope. General guidelines for doing so are provided.
\end{abstract}





\section{Introduction}

The technology of liquid mirrors (LMs) has been developed by the teams of Prof. Ermanno Borra at Laval University and Prof. Paul Hickson at the University of British Columbia (UBC). Early applications of the technology include the NASA Orbital Debris Observatory (NODO), a 3-m liquid-mirror telescope (LMT) used for observations of space debris (\citealt{paper11}) and several 3-m-class LMTs employed for lidar observations of the Earth's upper atmosphere (\citealt{paper12}, \citealt{paper13}). The largest LMT has been the 6-m Large Zenith Telescope (LZT), also used for lidar investigations (\citealt{paper2}). It was decommissioned in 2016. However, none of these LMTs was equipped with ad-hoc optical correctors for astrophysical observations and/or located at a good astronomical site.

A brief description of the International Liquid Mirror Telescope and its main components is given in Section 2. The unique characteristics of the ILMT, including the many advantages associated with zenith observations, are presented in Section 3. The ILMT sky survey consists mainly  of producing a large catalogue of astrometric and photometric measurements of all objects detected since the start of observations. This also made it possible to detect many interesting transients (see Section 4). At the BINA4 workshop, we presented a scale model of a liquid mirror telescope that can be used in school activities or public outreach. Section 5 contains brief instructions for students and young scientists to construct a model of a liquid mirror. The last section presents the conclusions. An illustrated overview of the ILMT project is available at the following URL: https://ilmt.be.  
 
\section{The ILMT and its main components}

\begin{figure}
	\centering
	\includegraphics[width=12.0cm]{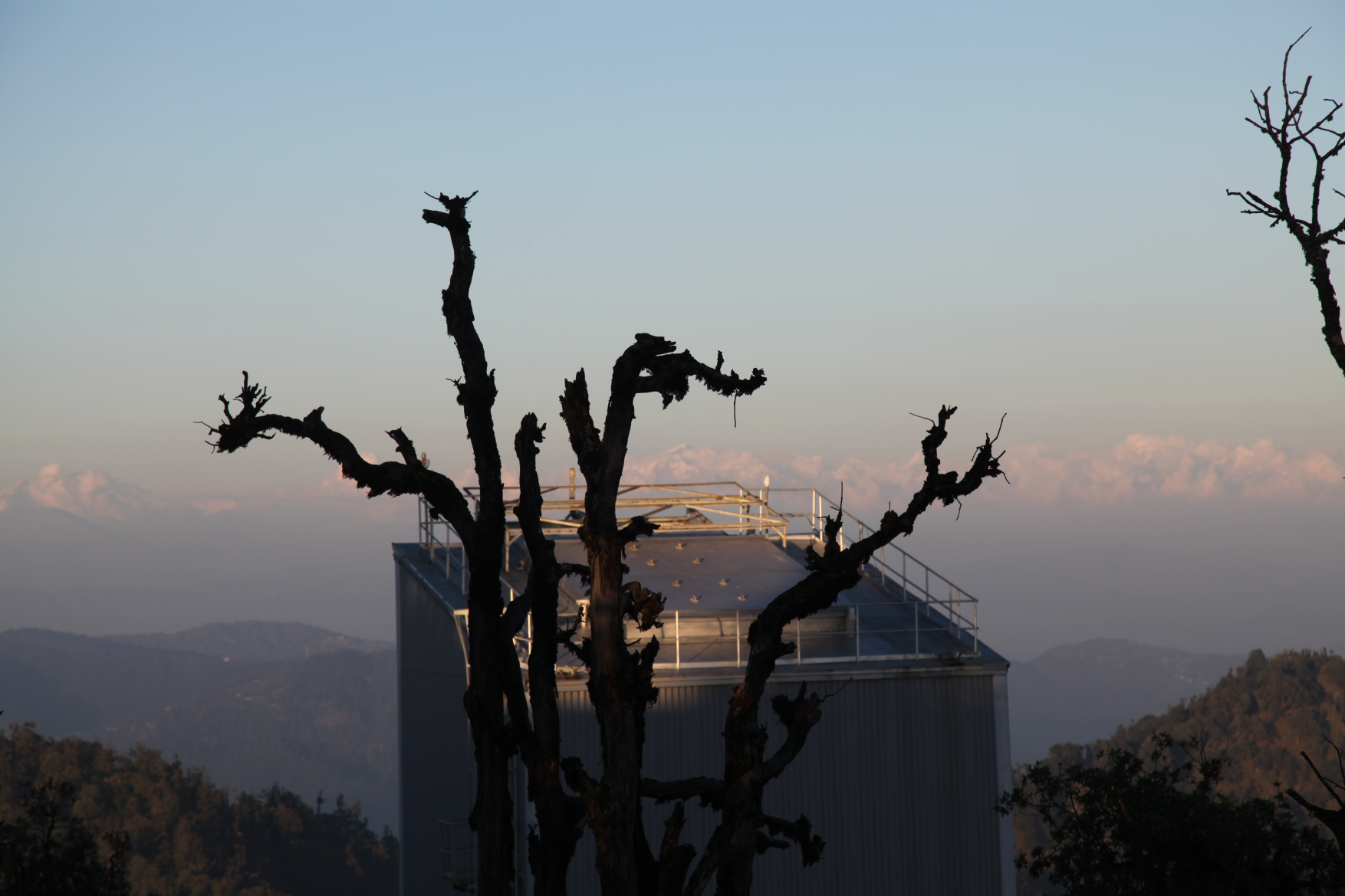}

	\begin{minipage}{120mm}
		\caption{The ILMT dome projected against the Himalayas. \label{ILMT_dome_5}}
	\end{minipage}
\end{figure}

A brief history of liquid mirror telescopes is available in  \citet{paper5} and \citet{paper19}. The ILMT was installed at the ARIES (Aryabhatta Research Institute of observational sciencES) Devasthal site, located in the foothills of the Himalayas (Uttarakhand, India, see Fig.~\ref{ILMT_dome_5} and \citealt{paper1,paper14,paper15,paper22,paper33}). 

\begin{figure}
	\centering
	\includegraphics[width=12.0cm]{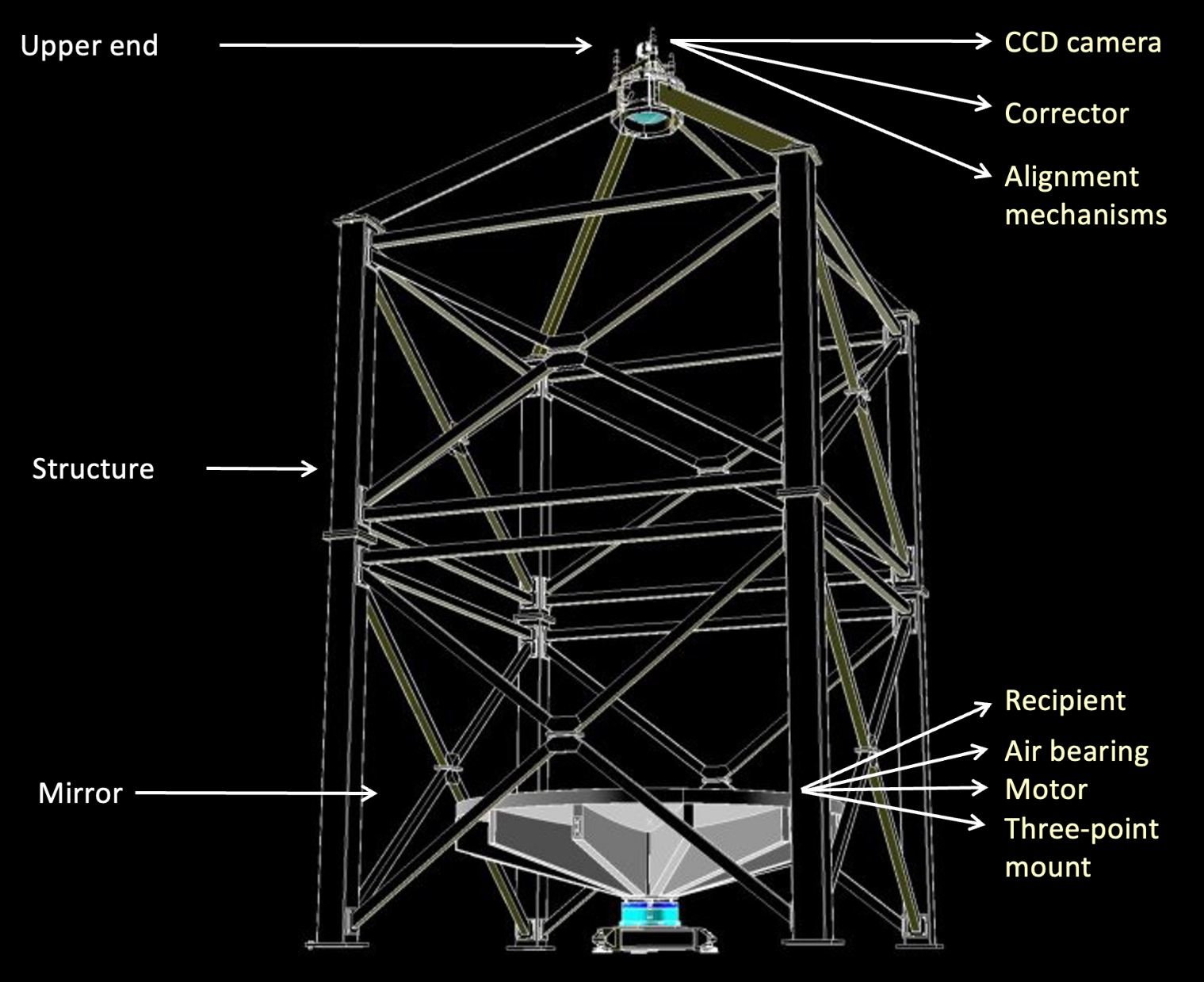}

	\begin{minipage}{120mm}
		\caption{The main components of the 4-m ILMT are indicated (see Section 2). \label{ILMT_components}}
	\end{minipage}
\end{figure}

The ILMT consists of a large paraboloid mirror, four meters in diameter,  mainly made up of a foam core sandwiched between two sheets of carbon fibre, slightly curved in a spherical shape. As it rotated at an angular speed of approximately 8 revolutions per minute, liquid polyurethane was poured inside. It solidified after half an hour, closely approximating (to within a few fractions of a millimetre) a paraboloid shape. A very thin layer (approximately 3 mm) of mercury was then pumped onto the mirror to reflect light from the sky. The total amount of mercury used is approximately 50 litres. The mirror is mounted on an air bearing to prevent the transmission of vibrations (see Fig.~\ref{ILMT_components}). An internal brushless DC motor keeps the angular rotation of the recipient constant. All this equipment rests on a three-point mount, to ensure that the optical axis of the telescope is parallel to the local axis of gravity. Since the telescope does not track the celestial objects, the trajectories of stars passing through the focal plane are branches of hyperbola. An optical corrector is therefore essential  to rectify these trajectories into straight lines (\citealt{paper4}). For more details on how the telescope works, see \citet{paper33}.


\begin{figure}
	\centering
	\includegraphics[width=12cm]{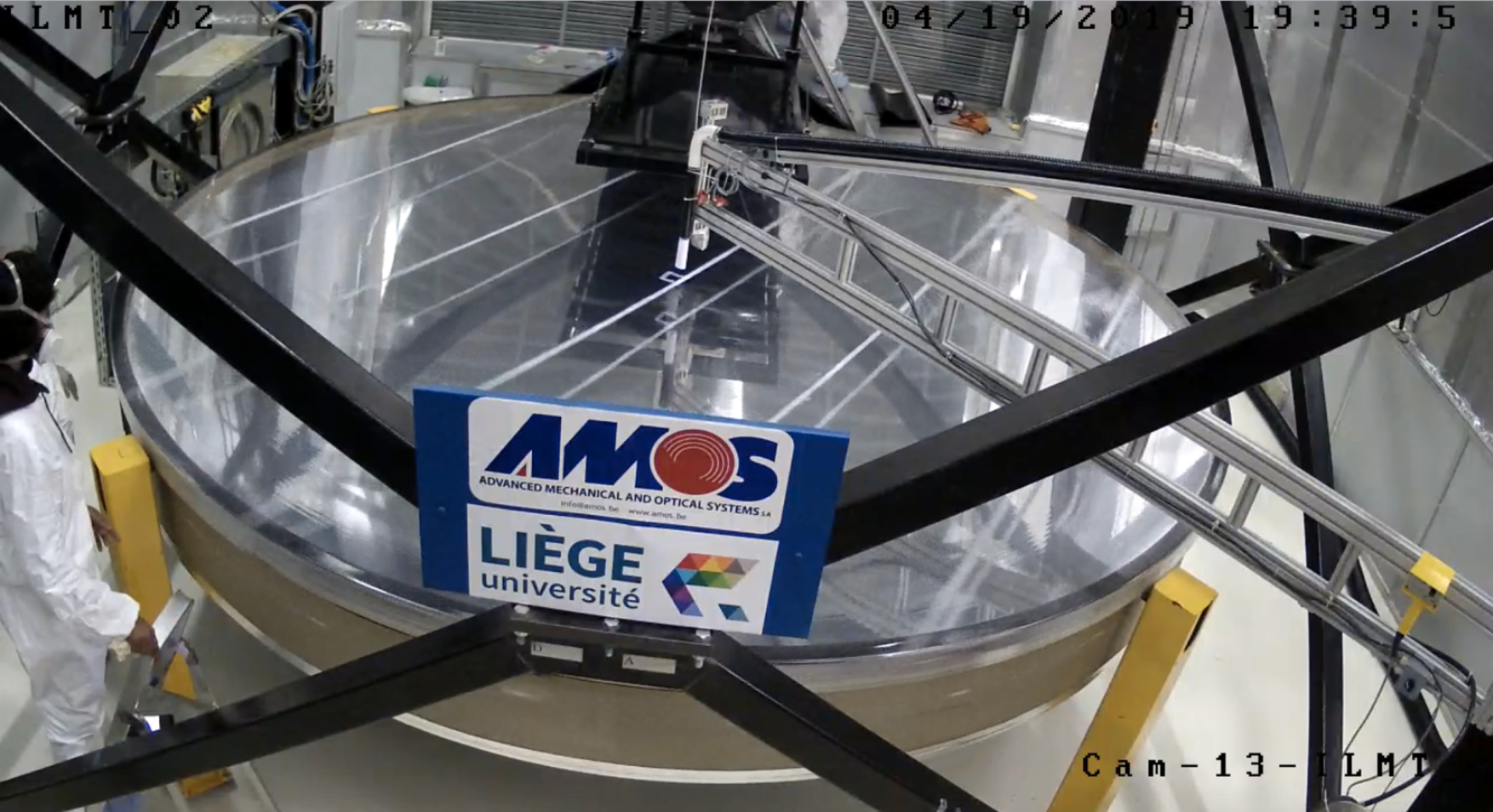}

	\begin{minipage}{120mm}
		\caption{Side view of the rotating 4-m mercury mirror covered with thin Mylar sheets.%
		\label{ILMT_1}}       
	\end{minipage}
\end{figure}

Due to the rotation of the Earth, the stars pass from one side of the detector to the other. This takes approximately 102 seconds. Thus, the electronic charges in the CCD are transferred from one row to the next at the same rate as the stars pass by. This mode of CCD operation is known as Time Delay Integration (TDI, \citealt{paper3}). The ILMT CCD camera is equipped with three Sloan broadband filters ($g', \, r' \, {\rm and } \, i'$, \citealt{Fukugita}). Fig.~\ref{ILMT_1} shows the mercury in the bowl and the thin Mylar sheets covering the mirror to prevent  friction between the mercury and the ambient air, which would cause ripples on the surface and affect image quality. It also helps prevent toxic mercury vapours from escaping into the ILMT enclosure.

\section{Unique features of the ILMT}

As the ILMT is a zenith-pointing telescope, it cannot move to follow celestial objects. The direct advantage is that the telescope always points to the best place in the sky where the seeing conditions and the image quality are optimal, atmospheric absorption is lowest and light pollution is least significant. In addition, observations are conducted continuously for approximately 17 minutes, after which each recorded image is transferred from the computer's main memory to the storage disk in less than a minute. Thus, if a single night of observation lasts 10 hours, we integrate light from the sky for approximately 9 hours and 25 minutes. The exposure time of $\simeq$ 17 minutes is essentially limited by the acquisition computer's RAM. A significant amount of overhead time is thus being saved thanks to the fixed pointing of the telescope.

Each single long CCD frame covers 1.21 square degrees and during a favourable night, we can record up to 30 such frames covering approximately 36 square degrees of the sky. Over the course of a  full year, we typically cover about 120 square degrees of the sky. Under good seeing conditions and dark sky, the limiting magnitudes easily reach 21.9, 21.7 and 21.5 in the Sloan spectral bands $g^\prime$, $r^\prime$ and $i^\prime$, respectively  (\citealt{paper43,paper40}). The ILMT is a unique optical survey telescope, the largest in India.

\section{The ILMT sky survey} 

The ILMT is a fast (f/2.36) optical telescope that observes the sky at the zenith in Devasthal (Latitude: $+29^\circ 
22^\prime 26^{\prime \prime}$) and thus recording images of all types of passing objects: stars, galaxies, quasars, asteroids, comets, extended nebulae, but also meteors, artificial satellites and space debris. 
All raw and calibrated data are freely accessible at 
\href{https://www.myqnapcloud.com/share/714efl5j416o5338t7yv1zzz_4d1gf282m0p226q5rsu39664c4cf21if#!/home}{ARIES arxiv} (\citealt{paper31}).

The size of the CCD is 4096 $\times$ 4096 pixels. The size and angular size of each pixel are $15~\mu$m and $0.3276^{\prime \prime}$, respectively. The width of the CCD therefore covers $22.3^\prime$ in declination. The gain and readout noise of the CCD are 3.9 e$^-$/ADU and 5 e$^-$ respectively. The characteristics of the ILMT CCD are described in detail in \citet{paper45}. 

\begin{figure}
	\centering
	\includegraphics[width=\textwidth]{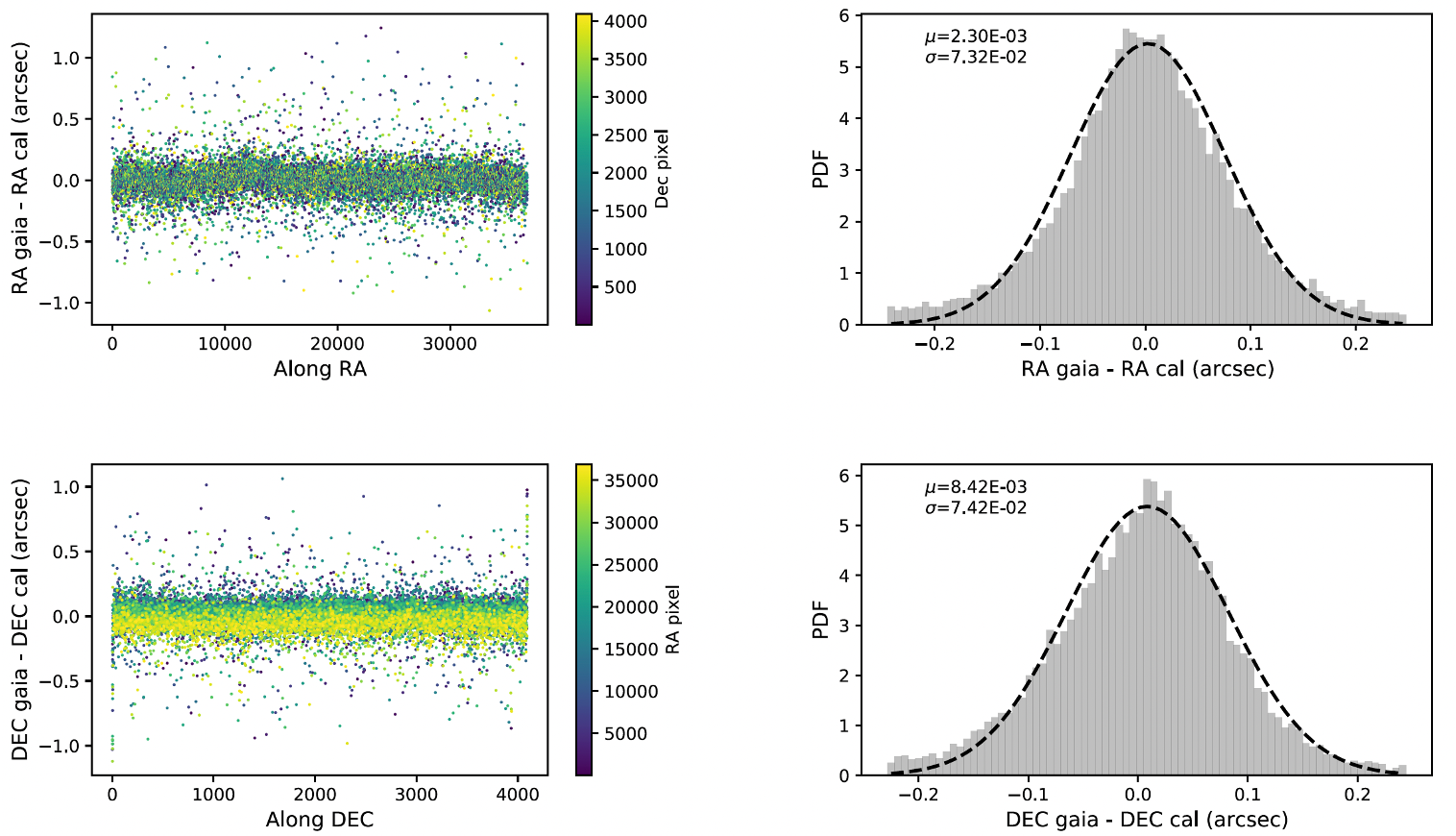}
	
	\begin{minipage}{120mm}
		\caption{\emph{Left:} Diagrams showing the differences between calibrated right ascension RA (resp.\ Declination Dec.) and Gaia RA (resp.\ Dec) of stars as a function of their RA (resp. Dec.) in a CCD frame measuring $9 \times 4096$ pixels, corresponding to a 17 min. long exposure.
		\emph{Right:}~Histograms of these differences indicating a mean quadratic astrometric dispersion of $\sim 0.07''$.%
		\label{ILMT_Astrometry}}
	\end{minipage}
\end{figure}

The astrometric accuracy of all detected celestial sources is typically better than one tenth of an arc second  (see Fig.~\ref{ILMT_Astrometry}, \citealt{paper27} and Negi et al. 2026 -- in preparation). When the ILMT field-of-view crosses the Milky Way, more than five million stars are being detected in a single night.
 
\begin{figure}
	\centering
	\includegraphics[width=\textwidth]{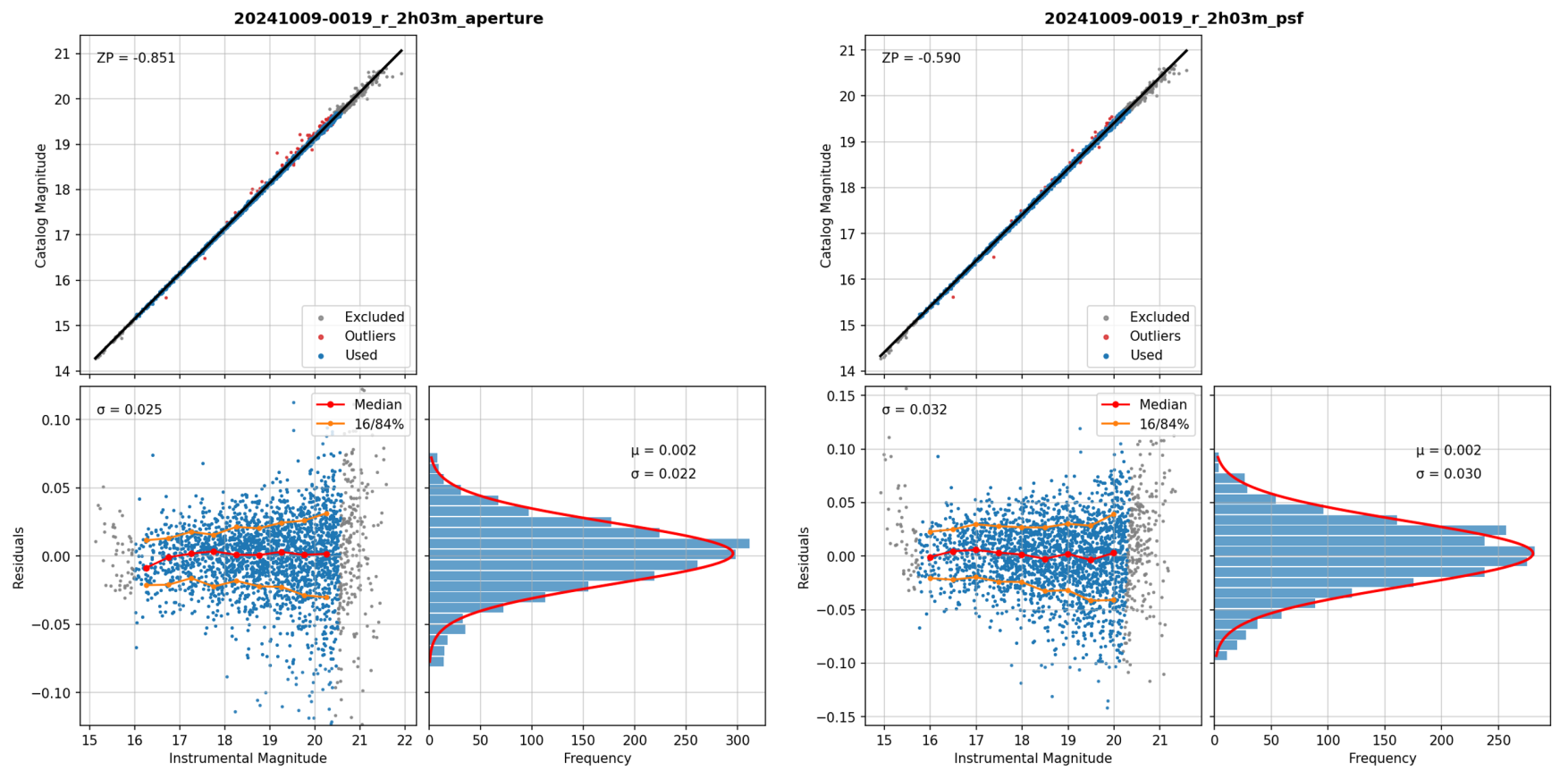}
	
	\begin{minipage}{120mm}
		\caption{%
			\emph{Left:} Aperture photometric calibration of the 19th CCD frame acquired on the night of 09-10-2024 at Local Sidereal Time 2h 11m with the $r'$ filter using ZTF photometric standard stars.
			\emph{Right:} PSF photometric calibration of the same frame. The uncertainties affecting these photometric calibrations are approximately 0.02--0.03\,mag.\ up to $r' = 21$.%
			\label{ILMT_Photometry}}       
	\end{minipage}
\end{figure}

With regard to the photometric calibration of the ILMT CCD frames, a preliminary catalogue of aperture and PSF photometric measurements is currently being compiled.  An example of the results obtained is shown in Fig.~\ref{ILMT_Photometry}. Further details are provided in \citet{paper28} and \citet{paper29}, and in these proceedings (\citealt{paper43,paper40}).

One of the main objectives of the ILMT team is to detect optical transients in near real time. The \texttt{PyLMT} pipeline that has been developed is based on image subtraction and uses a convolutional neural network for transient detection and classification (see \citealt{paper24,paper32,paper41,paper46}). 

\begin{figure}
	\centering
	\includegraphics[width=\textwidth]{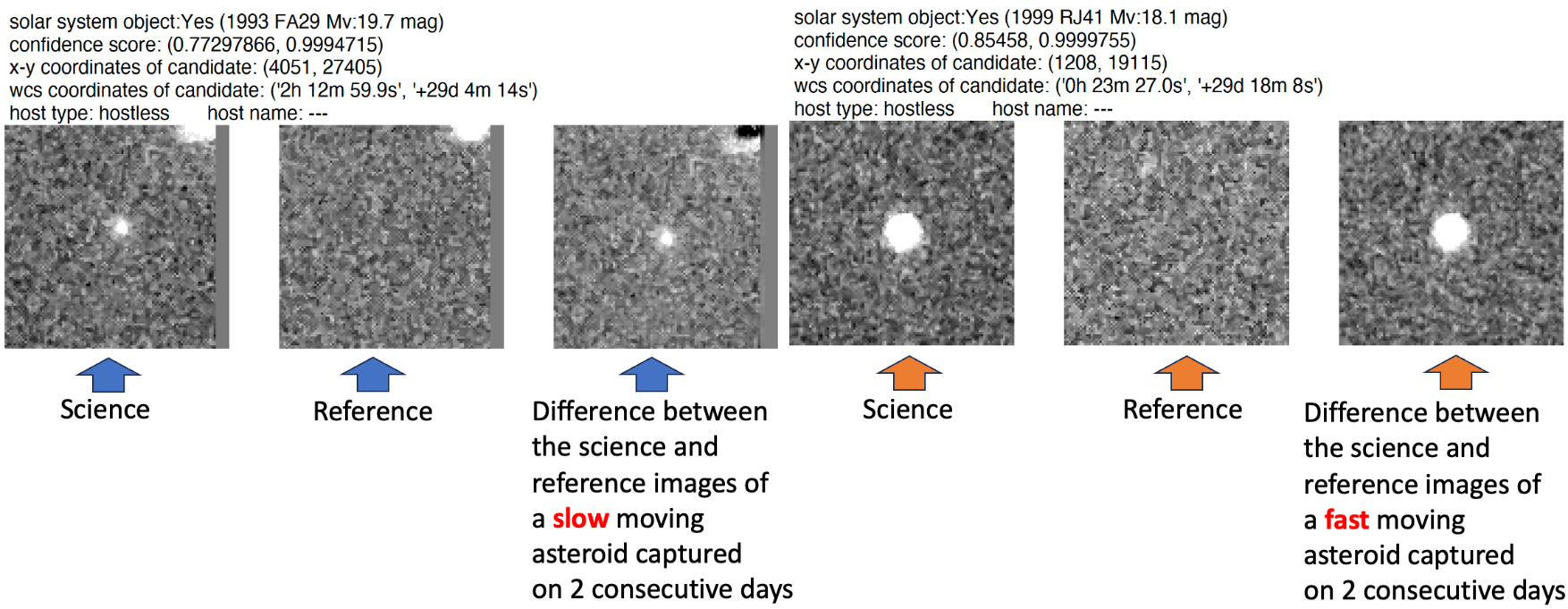}
	\vskip1ex
	
	\begin{minipage}{120mm}
		\caption{%
			Examples of detection with the ILMT of a slow-moving asteroid (3 images on the far left) and a fast-moving asteroid (3 images on the far right).
			Their actual detection appears on the science image.
			They are not visible on the reference image, and the result of subtracting the reference image from the science image clearly shows in both cases that the asteroid appears as a point-like object.
			The size of each image is approximately $30'' \times 30''$; North is up and East to the left.
			The slow-moving asteroid is characterised by a motion of $-8''$/hr and $-10''$/hr in right ascension and declination, respectively, while for the fast-moving asteroid, these figures are $\sim 63''$/hr and $29''$/hr.%
			\label{ILMT_MP}}       
	\end{minipage}
\end{figure}

Between November 2023 and May 2025, more than 20,000 transients associated with approximately 10,000 distinct objects were detected. Most of these transients are known minor planets (\citealt{paper25,paper39}). The ILMT is an excellent asteroid hunter thanks to its large aperture and short focal length. During the brief 102 sec. integration time, it is as if the image of each minor planet is frozen, even for the fastest moving asteroids (see Fig.~\ref{ILMT_MP}). Each asteroid image thus resembles a perfect PSF.

\begin{figure}
	\centering
	\includegraphics[width=\textwidth]{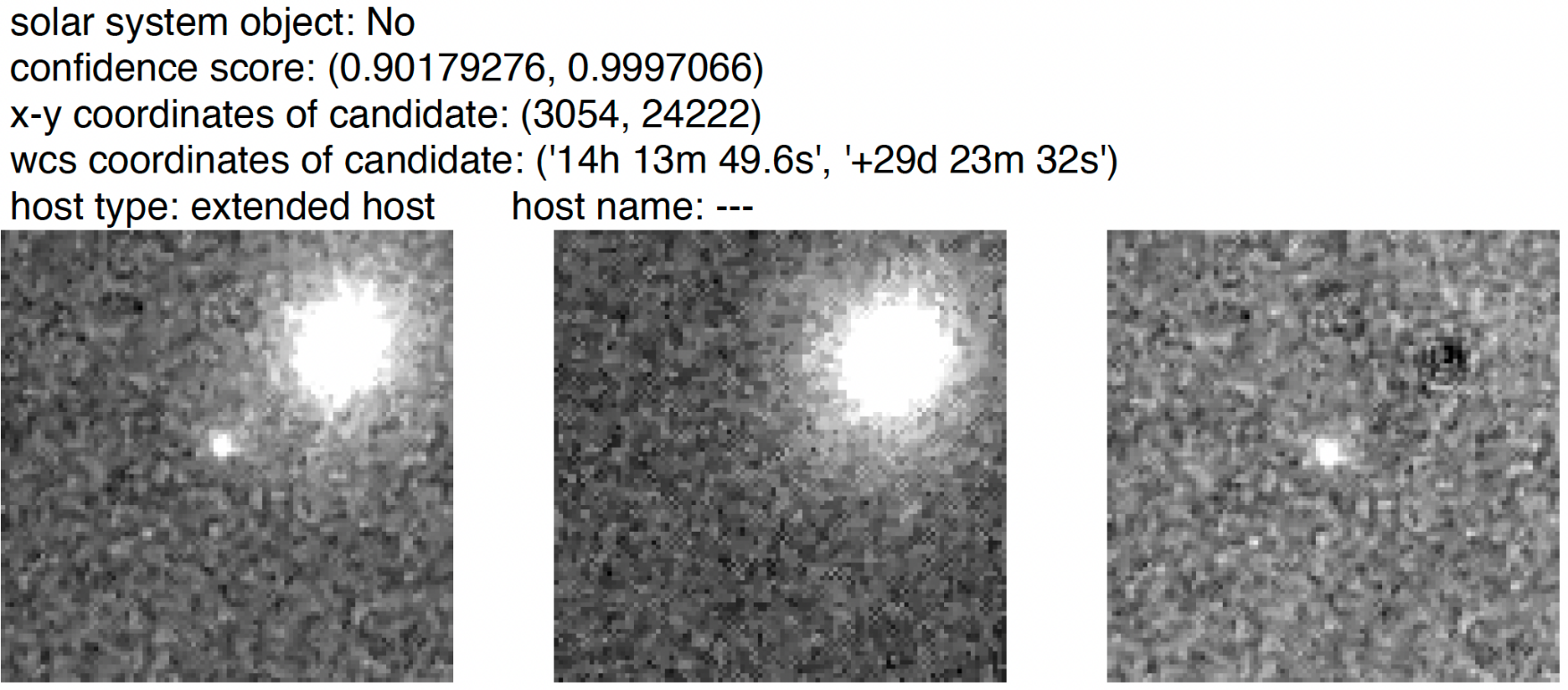}
	\vskip1ex
	
	\begin{minipage}{120mm}
		\caption{%
			\emph{Left:} Science image of the optical transient AT 2024fxn detected with the ILMT on 5/4/2024.
			\emph{Middle:}~Reference image of the same field observed with the ILMT on 15/2/2025.
			\emph{Right:}~Image resulting from the subtraction of the science image by the reference image. North is up and East to the left. The size of the images is $30'' \times 30''$.%
			\label{ILMT_SN}}       
	\end{minipage}
\end{figure}

Highly variable Active Galactic Nuclei (AGNs) are also detected (\citealt{paper26,paper37,paper47}), as well as low surface brightness galaxies (\citealt{paper36}) and interesting variable stars (\citealt{paper35}, Pranshu et al. 2026 -- in preparation). The most spectacular of these are supernovae (\citealt{paper7, paper8,paper23}). Since the ILMT began its observations, 22 supernovae have been detected, six of which are new discoveries by the ILMT team. Figure~\ref{ILMT_SN} illustrates the ILMT detection of one of them: AT 2024fxn on 5/4/2024. It was announced on the Transient Name Server (URL: https://www.wis-tns.org/object/2024fxn) on 7/4/2024 at 15:56:41 (UT) and confirmed by the ATLAS collaboration on the same day at 21:12:27, by the ZTF team on 8/4/2024 and by the Pan-STARRS survey on 17/6/2024. This illustrates the importance of rapid detection and announcement of detected optical transients.  Photometric and spectroscopic follow-up observations of supernovae detected with the ILMT are discussed in \citet{paper42} and \citet{paper44}.

\begin{figure}
	\centering
	\includegraphics[width=\textwidth]{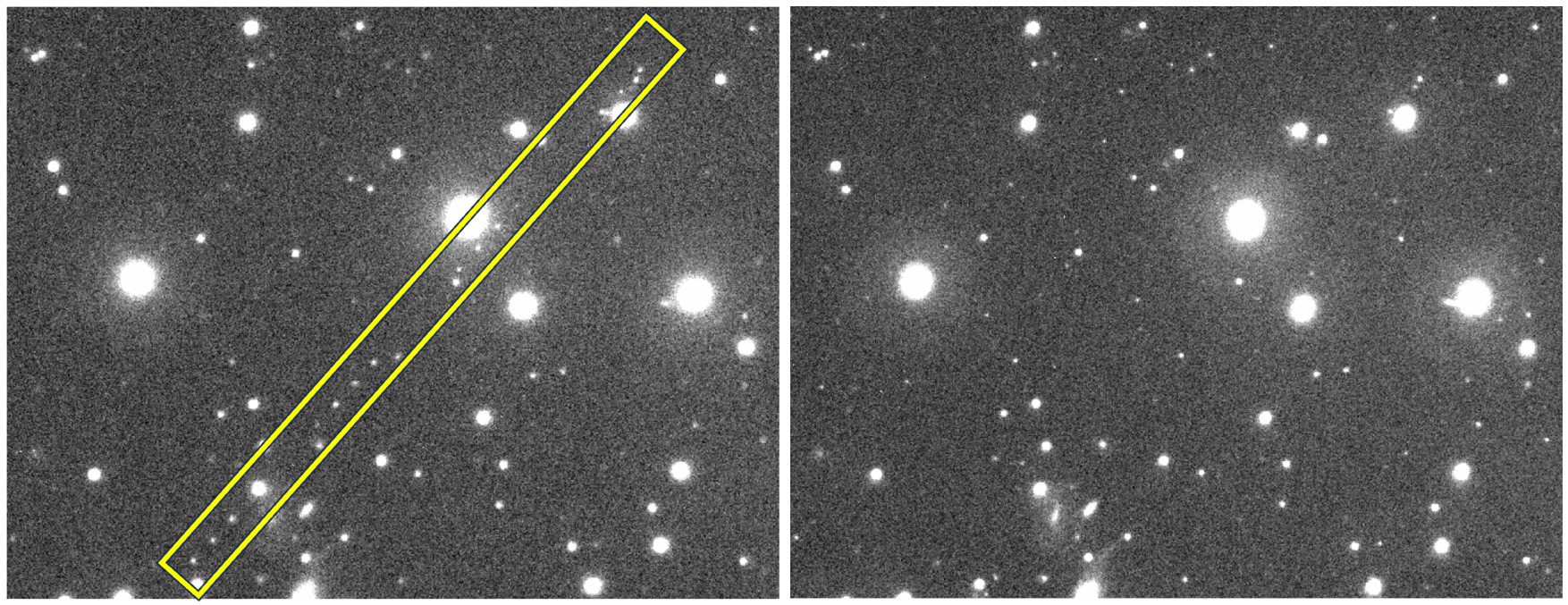}
	\vskip1ex
	
	\begin{minipage}{120mm}
		\caption{%
			\emph{Left:} Part of an ILMT CCD frame centred on RA(2000) = 23h42m13s, Dec(2000) = $+29^{\circ}\,04'\,26''$ on 2/11/2024 observed with the $i'$ filter.
			\emph{Right:}~Same part of the sky observed on 10/10/2024. Note the dozen or so of new point-like sources appearing inside the yellow rectangle. They are very likely caused by the specular reflection of a rotating space debris. The size of each field is approximately $6' \times 5'$. North is up and East to the left.%
			\label{ILMT_Debris}}       
	\end{minipage}
\end{figure}

Unfortunately, artificial satellites and space debris regularly pollute the  ILMT CCD frames (\citealt{paper30,paper34}, \citealt{paper38}). Specular reflection from some of these rotating space debris may sometimes produce star-like images on the ILMT CCD frames (see Fig.~\ref{ILMT_Debris}). Space debris  constitutes a real nuisance when performing direct imaging of the sky with the ILMT. The automatic detection and characterisation of space debris are discussed in \citet{paper38}.

\section{Scale model of a liquid mirror telescope}
\begin{figure}
	\centering
	\includegraphics[width=12cm]{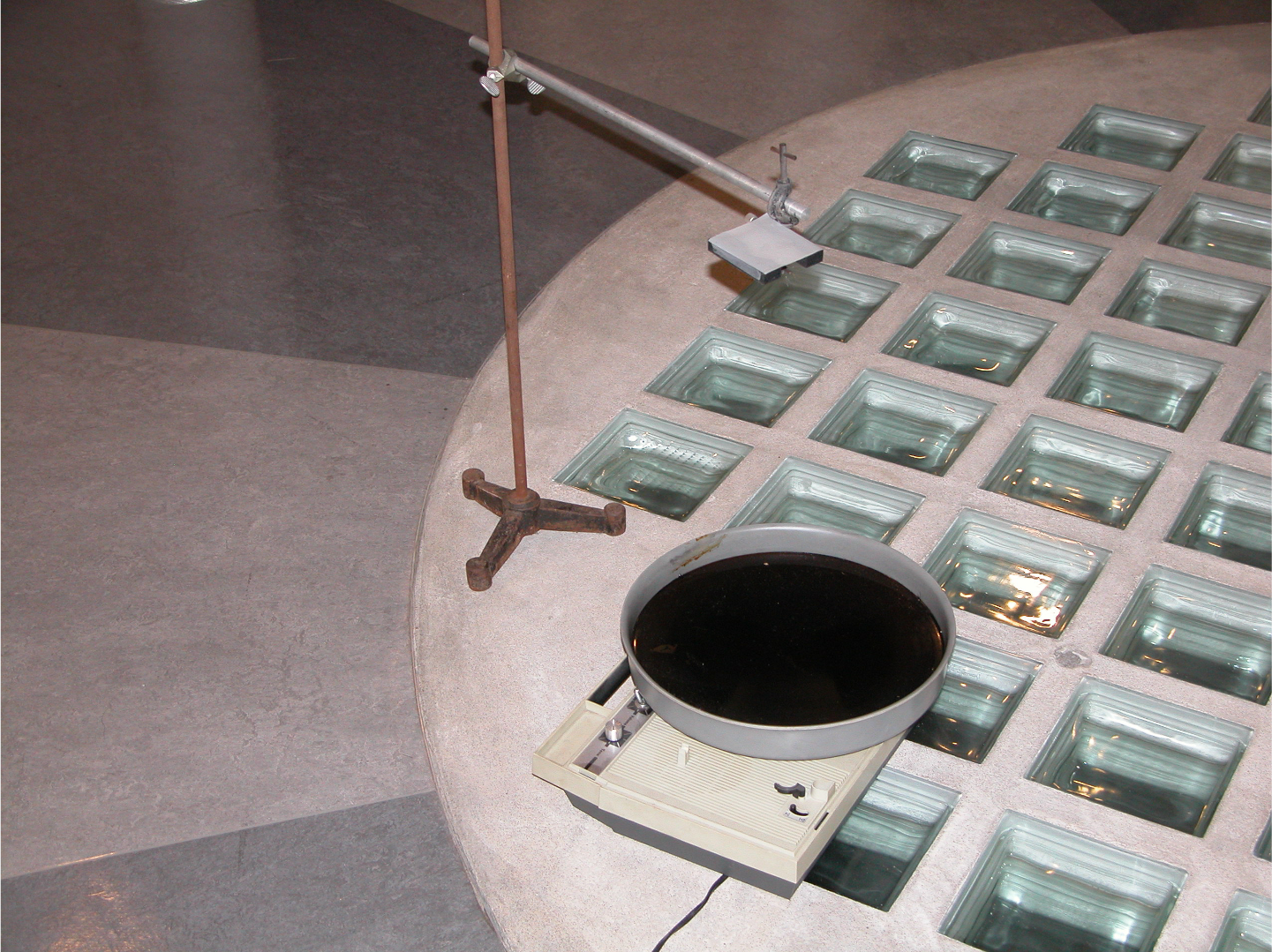}
	\vskip1ex
	
	\begin{minipage}{120mm}
		\caption{%
			Illustration of a pie tin filled with used motor oil, placed on a gramophone rotating at an angular speed of 33 revolutions per minute. The image of a light source placed approximately 4 metres above the mirror is (barely) visible on the baking paper placed in the focal plane.%
			\label{Gramophone}}       
	\end{minipage}
\end{figure}

We encourage students and young scientists to construct a scale model of a liquid mirror telescope. To do this, a flat reference plate is first needed, made of wood or metal, which can be placed horizontally using three adjustable bolts. Then it would be best to use an old gramophone (see Fig.~\ref{Gramophone}) or a turntable made by a mechanic and place it on the horizontal plate. A metal pie tin about 30 cm in diameter is also needed, which one can buy very cheaply. After placing the pie tin on the gramophone while it is rotating, for example at an angular speed of 33 revolutions per minute, some liquid resin (e.g. polyurethane) should be poured in. The resin will dry and solidify after about 30 minutes. The gramophone can now be stopped from rotating. A nice pre-paraboloid shape for the mirror should be obtained. Next, one can use some used car engine oil as a reflective liquid for the mirror. Let us simply pour it into the container and spin it again at the same angular speed. Using the relationship between the focal length $f$ of the paraboloid and the angular speed $\omega$ of the gramophone
\begin{equation}
  f = \frac{\omega^2}{2 \, g}, 
\end{equation} 
where $g$ represents the expression for the gravitational acceleration ($g \simeq 9.81$ m/s$^2$), the resulting focal length should be $f \simeq 60$ cm. The final result is a reflecting mirror with an aperture ratio $f/D \simeq 2$.

After installing a light bulb above the mirror, at a typical height of 3 - 4 meters or more, one can use a webcam or baking paper to find the image of the light source at a distance of $\simeq 60$~cm above the mirror. This experiment is part of public outreach activities organised by ARIES (Nainital, Uttarakhand, India).  

\section{Conclusions}
The ILMT is a zenith-pointing telescope entirely dedicated to a photometric and astrometric variability survey, as well as to the search for optical transients. Image quality is limited by the atmospheric turbulence. The CCD pixel size of $0.33^{\prime \prime}$ is well suited to the median local seeing ($\simeq 1.2^{\prime \prime}$). The ILMT field-of-view is $22.3^{\prime}\times22.3^{\prime}$. The liquid mirror telescope provides continuous zenith observations, characterised by the best image quality, which are also less affected by atmospheric extinction and light pollution. Public access to the data is free of charge at https://www.aries.res.in/facilities/astronomical-telescopes/ilmt.

Unfortunately, the ILMT CCD camera was found to be inoperative on 15 June 2025. It was sent for repair to Spectral Instruments in Tucson (Arizona, United States). 

An international contest will be organised in 2026 among young Indian scientists to design the best automatic search algorithm capable of detecting  and characterising as many and as faint as possible space debris in the ILMT data. 








\begin{acknowledgments}
The 4-m International Liquid Mirror Telescope (ILMT) project results from a collaboration between the Institute of Astrophysics and Geophysics (University of Li\`ege, Belgium), the Universities of British Columbia, Laval, Montreal, Toronto, Victoria and York, and Aryabhatta Research Institute of observational sciencES (ARIES, India). The team acknowledges the contributions of ARIES's technical and scientific staff. JS wishes to thank Service Public Wallonie, F.R.S.-FNRS (Belgium) and the University of Li\`ege, Belgium, for funding the construction of the ILMT. PH acknowledges financial support from the Natural Sciences and Engineering Research Council of Canada, RGPIN-2019-04369. PH, SF, AP-S and JS thank ARIES for hospitality during their visits to Devasthal. MD acknowledges the Innovation in Science Pursuit for Inspired Research (INSPIRE) fellowship award (DST/INSPIRE Fellowship/2020/IF200251) for this work. KP acknowledges the support from the Erasmus+ Programme of the European Union for a research visit to the Institute of Astrophysics and Geophysics, University of Li\`ege, Belgium.  TA thanks the Ministry of Higher Education, Science, and Innovations of Uzbekistan (grant FZ-20200929344). KM acknowledges the support from the BRICS grant DST/ICD/BRICS/Call-5/CoNMuTraMO/2023 (G) funded by the DST, India. JS and KM acknowledge the assistance received from the ANRF SERB-726 VAJRA Faculty Scheme, India. ARIES thanks the Department of Science and Technology (DST), Govt. of India, for the realisation of the project. Finally, we acknowledge the Belgo-Indian Network for Astronomy and Astrophysics (BINA) consortium for the local support to attend the 4th BINA workshop and for allowing us to present our work in the form of a paper in the proceedings of the fourth BINA workshop.
 \end{acknowledgments}

\begin{furtherinformation}

\begin{orcids}

\orcid{0000-0002-7005-1976}{Jean}{Surdej} 
\orcid{0000-0001-5115-6310}{Talat}{Akhunov}
 \orcid{0009-0000-1020-9711}{Bhavya}{Ailawadhi}
 \orcid{0009-0002-2621-6611}{Monalisa}{Dubey}
 \orcid{0000-0002-0394-6745}{Naveen}{Dukiya}
 \orcid{0000-0002-8289-8242}{Paul}{Hickson}
 \orcid{0000-0003-1637-267X}{Kuntal}{Misra} 
 \orcid{0000-0001-5824-1040}{Vibhore}{Negi} 
 \orcid{0009-0005-3754-2094}{Anna}{Pospieszalska-Surdej}
 \orcid{0009-0005-3844-3426}{Kumar}{Pranshu}
 
\end{orcids}

\begin{authorcontributions}
This work results from a long-term collaboration to which all authors have made significant contributions.
\end{authorcontributions}

\begin{conflictsofinterest}
The authors declare no conflict of interest.
\end{conflictsofinterest}

\end{furtherinformation}



%

\bibliographystyle{bullsrsl-en}

\bibliography{extra}

\end{document}